\documentclass[bibyear]{aa}  
\usepackage{natbib}
\usepackage{color}
\usepackage{graphicx}
\usepackage{txfonts}
\usepackage[draft]{hyperref}
\begin{document}

   \title{Machine learning in APOGEE:}

   \subtitle{Identification of stellar populations through chemical abundances}

        \author{
	Rafael Garcia-Dias\inst{1,2}
	\thanks{E-mail: rafaelagd@gmail.com}
	\and
	Carlos Allende Prieto \inst{1,2}
	\and
	Jorge S{\'a}nchez Almeida\inst{1,2}
	\and 
	Pedro Alonso Palicio \inst{1,2}
	}

	% List of institutions
	\institute{Instituto de Astrof\'{i}sica de Canarias,  E-38200 La Laguna,
		Tenerife, Spain\\
		\and
		Departamento de astrof\'{i}sica, Universidad de La Laguna,
		Tenerife, Spain\\
		}

   \date{Received February 06, 2019; accepted June 27, 2019}

% \abstract{}{}{}{}{} 
% 5 {} token are mandatory
	
	\abstract
	% context heading (optional)
	% {} leave it empty if necessary  
        {The vast volume of data generated by modern astronomical surveys offers test beds for the application of machine-learning. In these exploratory applications, it is important to {evaluate potential} existing tools and determine those that are optimal for extracting scientific knowledge from the available observations.}% Machine learning algorithms can be extremely valuable in the exploitation of the multidimensional data generated by spectroscopic surveys like the Apache Point Galactic Evolution Experiment (APOGEE).}
	% aims heading (mandatory)
	{We explore the possibility of using unsupervised clustering algorithms to separate stellar populations with distinct chemical patterns.}
	% methods heading (mandatory)
	{Star clusters are likely the most chemically homogeneous populations in the Galaxy, and therefore any practical approach to identifying distinct stellar populations should at least be able to separate clusters from each other. We have applied eight clustering algorithms combined with four dimensionality reduction strategies to automatically distinguish stellar clusters using chemical abundances of 13 elements. Our test-bed sample {includes} 18 stellar clusters with a total of 453 stars. }
	% results heading (mandatory)
	{We have applied statistical tests showing that some pairs of clusters {(}e.g., NGC 2458-NGC 2420{)} are indistinguishable from each other when chemical abundances from the Apache Point Galactic Evolution Experiment (APOGEE) are used. However, for most clusters we are able to automatically assign membership with {metric scores similar} to previous works. The confusion level of the automatically selected clusters is consistent with statistical tests that demonstrate the impossibility of perfectly distinguishing all the clusters from each other. These statistical tests and confusion levels establish a limit for the prospect of blindly identifying stars born in the same cluster based {solely} on chemical abundances.} %We also identify the chemical elements that are the most suitable for discriminating among families of stars with a common origin.}
	% conclusions heading (optional), leave it empty if necessary 
	{We find that some of the algorithms we explored are capable of blindly identify stellar populations with similar ages and chemical distributions in the APOGEE data. Even though we are not able to fully separate the clusters from each other, the main confusion arises from clusters with similar ages. Because some stellar clusters are chemically indistinguishable, {our study supports the notion of extending} {\textit{\textup{weak}}} chemical tagging that involves families of clusters instead of individual clusters.}
	

   \maketitle
   
    \newcommand{\besthomoscore}{0.807}
    \newcommand{\nelem}{13}
    \newcommand{\nclusters}{23}
    \newcommand{\ntsneelems}{13}
    \newcommand{\nstars}{453}
	\newcommand{\allelem}{Al, C, Ca, Fe, K, Mg, Na, N, Ni, O, P, S, and Si}
	\newcommand{\allabundances}{[Al/M], [C/M], [Ca/M], [Fe/H], [K/M], [Mg/M], [N/M], [Na/M], [Ni/M], [O/M], [P/M], [S/M], and [Si/M]}
    \newcommand{\Teff}{$T_\mathrm{eff}$}
	\newcommand{\LOGG}{$\log g $}
	\newcommand{\MH}{[M/H]}
	\newcommand{\CM}{[C/M]}
	\newcommand{\NM}{[N/M]}
	\newcommand{\aM}{[$\alpha$/M]}
	\newcommand{\Al}{[Al/H]}
	\newcommand{\C}{[C/H]}
	\newcommand{\Na}{[Na/H]}
	\newcommand{\N}{[N/H]}
	\newcommand{\V}{[V/H]}
	\newcommand{\Ca}{[Ca/H]}
	\newcommand{\Mg}{[Mg/H]}
	\newcommand{\Ox}{[O/H]}
	\newcommand{\Si}{[Si/H]}
	\newcommand{\Su}{[S/H]}
	\newcommand{\Ti}{[Ti/H]}
	\newcommand{\Fe}{[Fe/H]}
	\newcommand{\K}{[K/H]}
	\newcommand{\Mn}{[Mn/H]}
	\newcommand{\mTeff}{T_\mathrm{eff}}
	\newcommand{\mLOGG}{\log g}
	\newcommand{\mMH}{\mathrm{[M/H]}}
	\newcommand{\mCM}{\mathrm{[C/M]}}
	\newcommand{\mNM}{\mathrm{[N/M]}}
	\newcommand{\maM}{[\alpha/\mathrm{M}]}
	\newcommand{\mAl}{\mathrm{[Al/H]}}
	\newcommand{\mMg}{\mathrm{[Mg/H]}}
	\newcommand{\mFe}{\mathrm{[Fe/H]}}
	\newcommand{\mSi}{\mathrm{[Si/H]}}

%
%________________________________________________________________

\section{Introduction}

\quad Handling the overwhelming volume of data generated by many existing and forthcoming astronomical instruments is challenging. The future of astronomy depends on the development of efficient algorithms capable of making the best of all available data. Pattern recognition in high dimensional space is a fundamental task needed for the analysis of massive datasets. Spectroscopic surveys such as the Apache Point Galactic Evolution Experiment (APOGEE) \citep{majewski17}, allow two main approaches to target classification. It is possible to directly use the stellar spectra or to work with the corresponding atmospheric parameters and chemical abundances.

The first approach does not depend on stellar model atmospheres. At present, all the available models have limitations. They are either constrained by the lack of knowledge on the input atomic data (e.g., oscillation strengths) or are limited by the approximations made to accelerate the calculations, such as the assumption of plane-parallel geometry and local thermodynamic equilibrium. 

The legacy of the MK classification, named after W.W. Morgan and P.C. Keenan \citep{morgan43} to stellar astrophysics is undeniable. The classification is based on spectral features that are easily identifiable by visual inspection in medium-resolution spectra. This method does not depend on model atmospheres, but has the downside of being heavily supervised, which is inconvenient for the very large datasets.

The search for chemical abundance patterns relies on model atmospheres but enables a finer description of the stellar populations. While the spectral classification {is more sensitive to} physical parameters of the stars, i.e., effective temperature (\Teff) and surface gravity (\LOGG), pattern recognition in chemical abundance space can potentially uncover the star formation history of our Galaxy through the identification of chemical signatures shared by stars with a common origin, commonly referred to as {chemical tagging} \citep{freeman02}. We opt for the chemical abundances approach in this work, in order to test the feasibility of chemical tagging.  

Significant progress has been made in recent years using both strategies. For example, supervised spectral classification has been adopted in works such as those by \citet{Bailer-Jones1998, Singh, Bailer-Jones2001, Rodriguez2004, Giridhar, Manteiga2009}, and \citet{Navarro2012}. Unsupervised spectral classification was also explored in works such as \citet{sanchez09, vanderplas, sanchez10, Daniel2011, Morales-Luis11, sanchez13, sanchez16, Fernandez-Trincado2017, Matijevic2017, Price-Jones2017, Traven2017, Valentini2017, k-means, itamar}, and \citet{Price-Jones2019}.

The classification of stars based on chemical abundances has been explored by \cite{Blanco-Cuaresma2015} using homogeneous observations for 339 stars in 35 open clusters. Applying machine-learning techniques, \citet{Hogg2016a} were able to identify known star clusters and the Sagittarius dwarf galaxy in the APOGEE DR12 dataset. \cite{Schiavon2016} identified a stellar population that was unusually rich in nitrogen in the central part of the Galaxy. \citet{Kos2017} used GALactic Archaeology with HERMES (GALAH) data to spot new members of the Pleiades cluster. The application of machine-learning algorithms to stellar abundances was also employed by \citet{DaSilva2012, Ting2012, Jofre2017, Anders2018, Boesso2018}, and \citet{Price-Jones2019}.

We here explore the limits of a large battery of unsupervised clustering algorithms in distinguishing single-abundance-pattern stellar populations {in the chemical space explored by APOGEE}. Star clusters are likely the most chemically homogeneous populations in the Galaxy, thus any practical approach to chemical tagging should at least be able to separate known clusters from each other. Section 2 describes the details of the APOGEE data used in this work. Section 3 presents the star cluster samples. Section 4 discusses the feasibility of the task. Section 5 presents the algorithms we tested, and Section 6 describes the results. Finally, Section 7 summarizes the work and discusses the viability of applying these algorithms to blindly identifying families of stellar populations in the complete APOGEE dataset. {We remark that all conclusions derived from the result presented in this study are restricted to the stellar population and chemical elements we explored. The APOGEE abundances from IR spectra are restricted to the alpha-capture elements and Fe-peak elements. The dimensionality of the APOGEE's elemental abundance space is expected to be restricted. Including other elements can potentially change the results we found here.} 

\section{Data}

\quad APOGEE is spectroscopically observing hundreds of thousands of stars, primarily giants, which emphasizes regions that are obscured by dust: the Galactic plane and the central parts of the Galaxy. With these observations, a chemical map is created \citep{blanton17, majewski17}. The project employs two twin spectrograph on 2.5m telescopes \citep{gunn06} on the Northern (using the Sloan Digital Sky Survey, SDSS telescope at Apache Point Observatory, USA) and Southern (using the DuPont Telescope at Las Campanas, Chile) hemispheres. This work is based on data from SDSS Data Release 14 \citep{DR14}, which includes observations of more than 260,000 stars.

With a signal-to-noise ratio per half a resolution element $>100$, the $H-$band APOGEE spectra, analyzed by the APOGEE Stellar Parameters and Chemical Abundances Pipeline \citep{ana16}, provide determinations of the chemical abundances for about 20 elements with a precision that ranges from 0.03-0.05 dex for iron, oxygen or magnesium in solar metallicity K-giants to $\sim$ 0.2 dex for nitrogen in moderately metal-poor stars \citep{Holtzman15, Bertran16, bovy16}. Of these elements, we retained those that have been shown to be reliable in previous studies (\citealt{Holtzman15, holtzman18}; \citealt{herik18}). The elements used in the clustering applications are \allelem.

APOGEE includes observations of globular clusters and many open clusters targeted for science and calibration purposes (see, e.g., \citealt{frinchaboy13}). Most clusters have a fairly limited extent on the sky and pose a challenge for the 70-arcsecond fiber collision radius of APOGEE. Only a limited number of members are therefore typically observed per cluster. Only a few of the clusters targeted by APOGEE have tens of observed stars (e.g., NGC 2420, NGC 6791, the Pleiades, or M67; see, e.g. \citealt{cunha15}, \citealt{linden17}; \citealt{diogo16, diogo18}). About 20 clusters include 5--30 stars each, and those are the ones we focus on in this paper. APOGEE provides the abundances determined directly from the fitting of the observations with synthetic spectra based on standard model atmospheres. The optimization of the parameters is based on a $\chi^2$ criterion, and it is performed using the FERRE code\footnote{FERRE is available from github.com/callendeprieto/ferre} \citep{ferreIII}. These abundances show mild trends with the stellar effective temperature for star members of open clusters. Even though diffusion in stellar envelopes can systematically alter photospheric abundances in a manner that depends on stellar mass (see, e.g., \citealt{diffusion, diogo18}), and therefore on effective temperature, such trends are interpreted mainly as a result of shortcomings in the model atmospheres and radiative transfer calculations that are used to derive the abundances, for example, departures from local thermodynamic equilibrium (LTE) or 3D effects. In an attempt to remove such effects, the trends with effective temperature are modeled with smooth functions, that take all the open clusters into account. This produces internally calibrated abundances that are the default abundances that are published in the data releases.

We are interested in relative abundances of stars with similar atmospheric parameters and not in absolute values, therefore we preferred to use the { uncalibrated} abundances. These are released together with the calibrated abundances for each data release\footnote{See the data model and specifically the description of the FELEM array in http://www.sdss.org/dr14/irspec/abundances/}. Far more stars have uncalibrated abundances than the calibrated ones because the calibrations span a limited range in parameters space. {We did not use dwarfs in this particular study, but dwarf stars have only uncalibrated values, and they make up about 20\% of the total APOGEE sample. Most importantly,} the smooth functions used to correct the trends with effective temperature may lose some of the precision of the original uncalibrated values for stars with similar parameters. {In addition, the calibration process requires placing the abundances for all clusters on the same scale, and therefore adopting a fiducial metallicity for each cluster, which may introduce additional offsets. Some of the elements, such as Na, P and S, present offsets compared with abundance results from spectroscopy in the visible. In this work, however, we explore the relative differences among the clusters that are not affected by these issues. From a practical standpoint, we decided to test our analysis on both calibrated and uncalibrated abundances. We found better\footnote{Clusters were more clearly distinguished from each other.} results with the latter. {In Appendix \ref{sec:ap} we present the results for the same analysis as described in Sections \ref{sec:sample}, \ref{sec:dist}, \ref{sec:clustering} and \ref{sec:results}, but for the results in the appendix we used the calibrated abundances.}

\section{Sample}
\label{sec:sample}
\quad We selected clusters in APOGEE DR14 with at least five members. Membership was determined based on radial velocities and on the distribution of chemical abundances. We selected the stars that were compatible with the mean radial velocities and velocity dispersions in \citet{dias} for open clusters and those in \citet{francis} for globular clusters. We then applied one iteration of two-$\sigma$ clipping in all the chemical abundances to guarantee a single composition. Table \ref{tab: clusters} shows the list of selected clusters. We provide an online table with a list of all the stars in the clusters\footnote{\url{http://cdsarc.u-strasbg.fr/ftp/vizier.submit//Garcia-Dias/Table1.csv}}. Figure \ref{fig:HR_diagram} shows the distribution of \Teff\, and \LOGG\, for the whole sample.

\begin{figure*}
	\centering
	\includegraphics[width=0.95\textwidth]{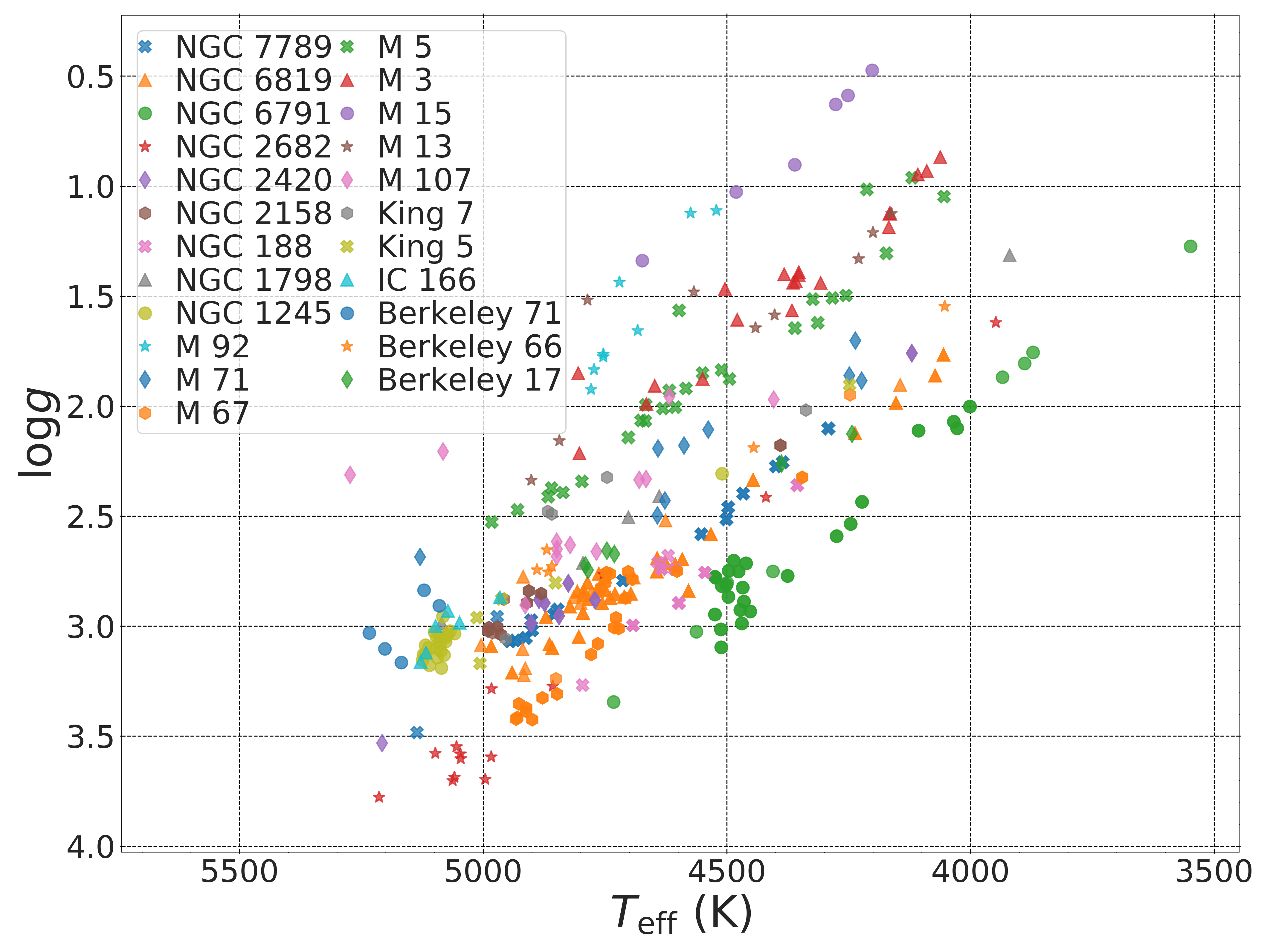}
	\caption{\label{fig:HR_diagram} Distribution of \Teff\, and \LOGG\, for all the stars used in the unsupervised clustering exercise. For each cluster we used a unique combination of symbol and color, as shown in the plot legend.}
\end{figure*}

\begin{table}
	\begin{center}
		\caption{\label{tab: clusters} Clusters used in this study.}
		\begin{tabular}{lccc}
			Cluster   &  N$_{\star}^{i}$ & $\log$(Age)$^{ii}$ (years) & $\langle{\mathrm{[Fe/H]}}\rangle_{\mathrm{uncalib}}^{iii}$ \\ \hline
			\multicolumn{4}{c}{\textbf{Open Clusters}}                   \\
	        \object{King 7}      &         5 &    8.82$^{\mathrm{iv}}$   &   -0.06 \\
			\object{Berkeley 71} &         5 &    9.00$^{\mathrm{iv}}$   &   -0.21 \\
			\object{IC 166}      &         6 &    9.00$^{\mathrm{iv}}$   &   -0.07 \\
			\object{NGC 2158}    &        18 &    9.02$^{\mathrm{iv}}$   &   -0.15 \\
			\object{NGC 1245}    &        22 &    9.03$^{\mathrm{iv}}$   &   -0.06 \\
			\object{King 5}      &         6 &    9.10$^{\mathrm{iv}}$   &   -0.12 \\
			\object{NGC 7789}    &        32 &    9.15$^{\mathrm{iv}}$   &    0.03 \\
			\object{NGC 1798}    &         5 &    9.25$^{\mathrm{iv}}$   &   -0.19 \\
			\object{NGC 2420}    &        15 &    9.30$^{\mathrm{iv}}$   &   -0.13 \\
			\object{NGC 6819}    &        81 &    9.36$^{\mathrm{iv}}$   &    0.10 \\
			\object{NGC 2682}    &        14 &    9.45$^{\mathrm{iv}}$   &    0.05 \\
			\object{M 67}        &        46 &    9.45$^{\mathrm{iv}}$   &    0.07 \\
			\object{Berkeley 66} &         6 &    9.60$^{\mathrm{iv}}$   &   -0.15 \\
			\object{NGC 188}     &        15 &    9.88$^{\mathrm{iv}}$   &    0.13 \\
			\object{NGC 6791}    &        58 &    9.92$^{\mathrm{iv}}$   &    0.39 \\
			\object{Berkeley 17} &         6 &   10.00$^{\mathrm{iv}}$   &   -0      .12 \\
			\multicolumn{4}{c}{\textbf{Globular Clusters}}               \\
			\object{M 5}         &        27 &   10.62$^{\mathrm{v}}$  &   -1.17 \\
			\object{M 3}         &        20 &   11.39$^{\mathrm{v}}$  &   -1.38 \\
			\object{M 13}        &         9 &   11.65$^{\mathrm{v}}$  &   -1.49 \\
			\object{M 15}        &         6 &   12.93$^{\mathrm{v}}$  &   -2.39 \\
			\object{M 71}        &         9 &   13.70$^{\mathrm{v}}$  &   -0.71 \\
			\object{M 107}       &        12 &   13.95$^{\mathrm{v}}$  &   -0.97 \\
			\object{M 92}        &         8 &   14.20$^{\mathrm{vi}}$ &   -2.33 \\
			All                 &           431 &   --                                 &   --    \\ \hline
			\multicolumn{4}{l}{\scriptsize{i: N$_{\star}$ is the number of cluster members.}} \\ 
			\multicolumn{4}{l}{\scriptsize{ii: The logarithm of the cluster ages.}} \\ 
			\multicolumn{4}{l}{\scriptsize{iii: Mean iron abundance.}} \\ 
			\multicolumn{4}{l}{\scriptsize{iv: \cite{dias}}} \\ 
			\multicolumn{4}{l}{\scriptsize{v: \cite{gc1}, \cite{gc2}}} \\ 
			\multicolumn{4}{l}{\scriptsize{vi: \cite{gc3}}}
		\end{tabular}
	\end{center}
\end{table}

\section{Cluster distinguishability according to chemical abundance}
\label{sec:dist}

\quad In order to determine whether the clusters can be distinguished from each other based on their abundances distributions, we have performed two statistical tests. The 1D Kolmogorov-Smirnov two-sample test (K-S test; \citealt{smirnov39, darling57}) was applied for each pair of clusters and each chemical element, and the Cramer multivariate test \citep{Baringhaus2004, Elias, Yeremi2014} was applied for each pair of clusters and then for all chemical elements at once. In both tests, the outcome is the probability (p-value) that the two samples are drawn from the same distribution (the null hypothesis). If the p-value is lower than 0.01, we can reject the null hypothesis with a confidence level of 99\%. That is to say, we can conclude that the two samples are unlikely to be drawn from the same distribution, and the clusters can be considered distinguishable. \footnote{We used the \href{https://www.scipy.org/}{SciPy} software library version 1.1.0 to perform the K-S tests.}

We first applied the tests to subsamples of stars in each individual cluster in order to understand the systematics of the tests. For this, we took the median p-value over 1000 tests performed with random subsamples of each cluster. The subsamples were obtained by dividing the cluster into two groups of nearly equal size. Figure \ref{fig: C} shows the results of applying the K-S test for carbon. We note that Figure \ref{fig: C} is symmetric with respect to the main diagonal. We exploit this symmetry in Figures \ref{fig: ks-test} and \ref{fig: cramer} to facilitate the direct comparison of the outcomes of different tests. The main diagonal presents the median p-value of the 1000 subsamples of each cluster, and the off-diagonal cells give the result for a single run using all stars in each pair of clusters. This figure shows that it is not possible to distinguish many pairs of cluster based on their carbon abundances alone. For instance, M3 and M13 have indistinguishable carbon distributions. In some cases, even globular and open clusters have indistinguishable carbon distributions, as is the case for the pair M67-M71.

\begin{figure*}
	\centering
	\includegraphics[width=\textwidth]{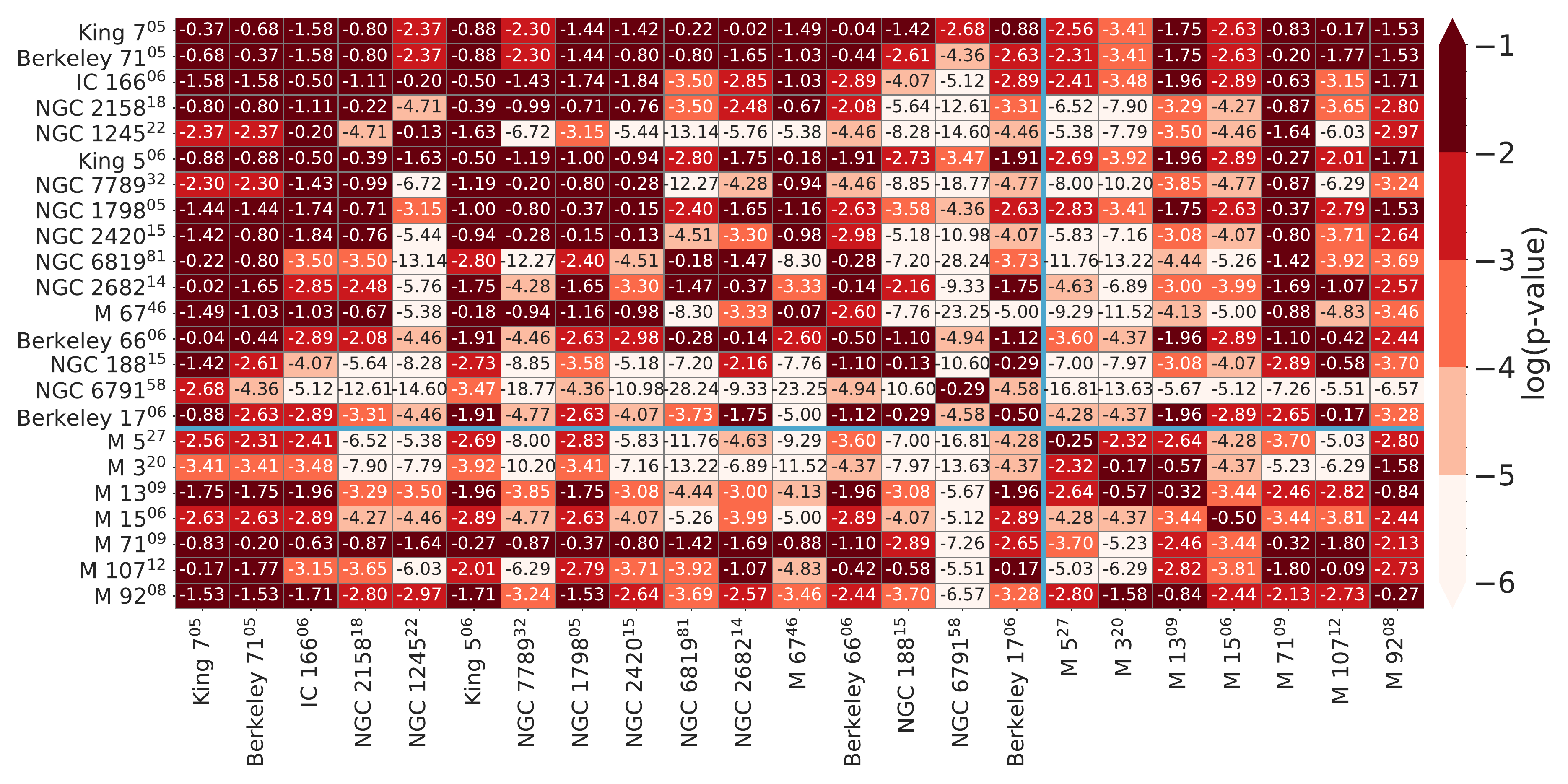}
	\caption{\label{fig: C} Logarithm of the p-values of the K-S two-sample test for the carbon distributions of each pair of clusters. The cells are colored in five shades of red, from light to dark in logarithmic scale. Inside each cell, we show the p-value of the test for that particular pair of clusters. Superscripts in the cluster names indicate the number of stars in the cluster. The main diagonal shows the median p-value of 1000 tests resulting from randomly dividing the cluster into two subsamples of nearly equal sizes. Two blue lines separate the objects into globular clusters (lower right) and open clusters (upper left). Dark red represents values greater than 0.01: we cannot reject the null hypothesis for these cases.}
\end{figure*}

In contrast, performing the K-S test for all pairs of clusters and taking the lowest p-value of all the \nelem\, elements, we show that, except for the pair Berkeley 17-Berkeley 71, all the \nclusters\, clusters can be distinguished from each other in at least one element, as shown in Figure \ref{fig: ks-test}. The main diagonal in the figure is calculated as explained above for Figure \ref{fig: C}, but showing only the lowest p-value of all the elements. In Figure \ref{fig: ks-test} we combine the results for C with those obtained for all elements. All the cells above the diagonal represent the lowest p-value of all elements to that pair of clusters. The cells below the main diagonal represent the results for carbon. In each cell, we identify the element for which the p-value is presented.

\begin{figure*}
	\centering
	\includegraphics[width=0.995\textwidth]{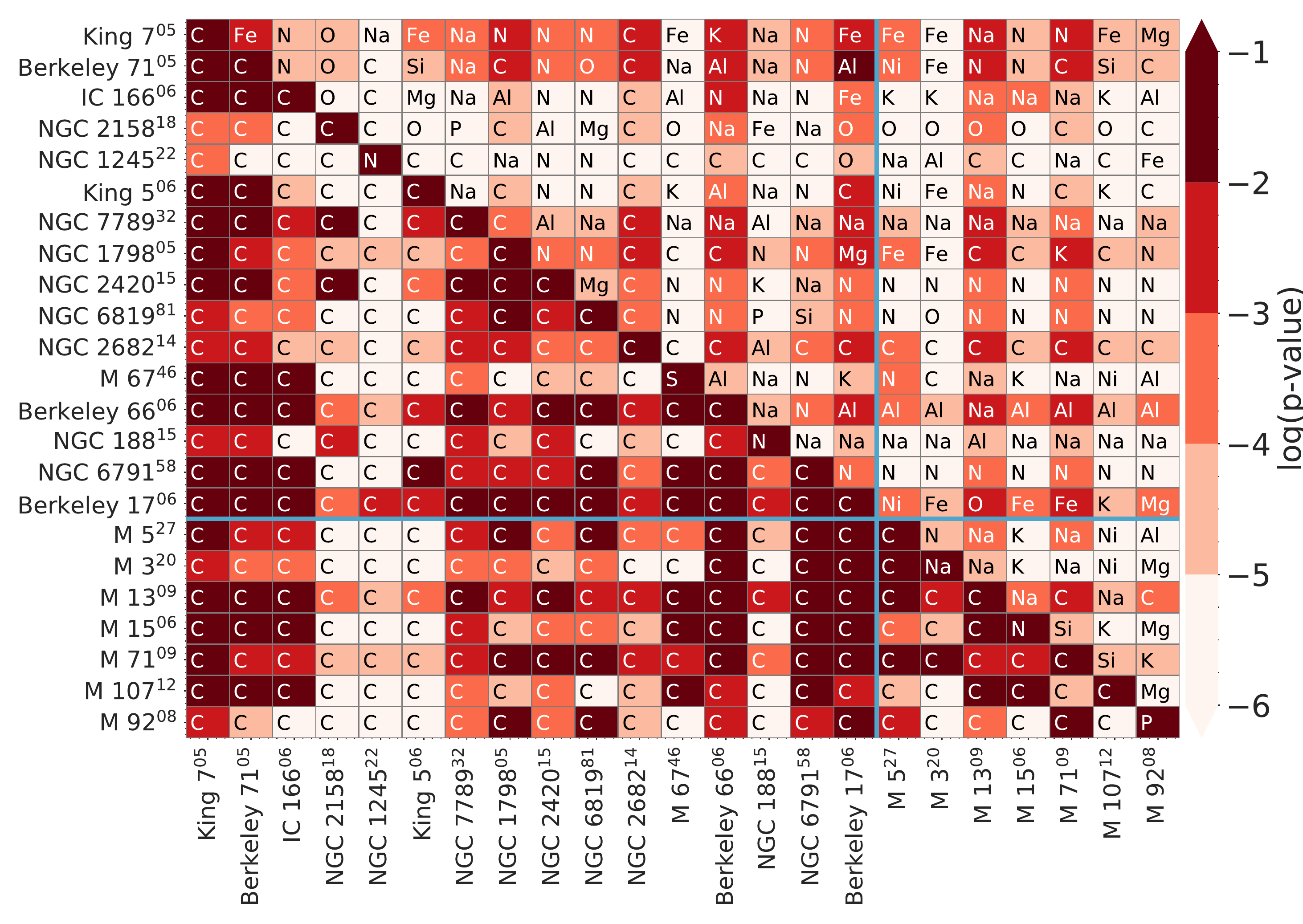}
	\caption{\label{fig: ks-test} Logarithm of the p-values of the K-S two-sample test for each pair of clusters. The elements of the main diagonal and above the main diagonal represent the lowest p-value found among all the elements, and while the elements below the main diagonal represent the p-value for carbon. The cells are colored in shades of red as shown in the color bar. Red colors are saturated at 0.01. Superscripts in the cluster names indicate the number of stars in the cluster. At the center of each cell, we show the element for which the p-value is calculated. Two blue lines separate the objects into globular and open clusters.}
\end{figure*}

Three main conclusions can be drawn from Figure \ref{fig: ks-test}: (1) the probability stemming from the K-S test is extremely low for many of the pairs; it is lower than $10^{-5}$ in many cases. The clusters can therefore be distinguished in at least one element. P-values higher than $10^{-3}$ are found for several pairs, for example, M13-M71 and NGC 7789-NGC 2682, and therefore we expect that it will be harder to separate these pairs by any unsupervised clustering algorithm than it would be for pairs with p-values lower than $10^{-5}$. (2) All clusters present high median p-values when compared with themselves. This fact underlines the cohesion of the clusters and demonstrates the consistency of the test. (3) A few elements are particularly important for distinguishing the clusters from each other.

Table \ref{tab:best_elem} contains the times that each element is best for distinguishing a unique pair of clusters. Five elements are sufficient to separate 80\% of the pairs. Calcium never appears to be the best element to distinguish a pair of clusters. In addition, the most useful elements for distinguishing pairs of globular clusters are not those that are most useful for separating pairs of open clusters, or pairs composed of a globular and an open cluster. We see that Na and C are particularly well suited to separate globular clusters from each other, and N, Na, and C are excellent in separating pairs of globular and open clusters. Fe might be expected to be very effective in distinguishing globular and open clusters from each other, but this element often presents higher p-values in the K-S test when pairs of distributions are compared that containing open and globular clusters than the p-value we found for other elements. Pairs of open clusters are best distinguished using C, N, and Na.

The results of the Cramer test are presented in Figure \ref{fig: cramer}. The main diagonal is also the median value of 1000 runs of the test over random subsamples of the clusters, as described for the K-S test. The Cramer test has a stochastic nature, and thus some variation is expected. In order to avoid picking an atypical outcome, the values presented in the cells below the main diagonal are the median value of 100 runs of the Cramer test. The cells above the main diagonal represent the lowest p-values of the K-S test. %For all pairs with a p-value lower than 0.01, the standard deviation over the 100 runs is lower than 0.003.

Figure \ref{fig: cramer} shows that we cannot reject the null hypothesis for ten pairs of clusters: Berkeley 71-King 7, IC 166-King 7, NGC 1245-King 7, King 5-King 7, Berkeley 66-King 7, Berkeley 17-King 7, King 5-Berkeley 71, Berkeley 66-Berkeley 71, Berkeley 17-Berkeley 66 and M15-M92. In these ten pairs, only six have p-values higher than 10$^{-3}$ in the K-S test. This discrepancy between the tests is expected, because the problem of comparing multidimensional samples is not trivially reduced to 1D pieces as we attempt to do in the exercise with the K-S test. However, we present the analysis with the K-S test because it is intuitive and provides insight on which chemical elements better distinguish the clusters from each other.

\begin{figure*}
	\centering
	\includegraphics[width=\textwidth]{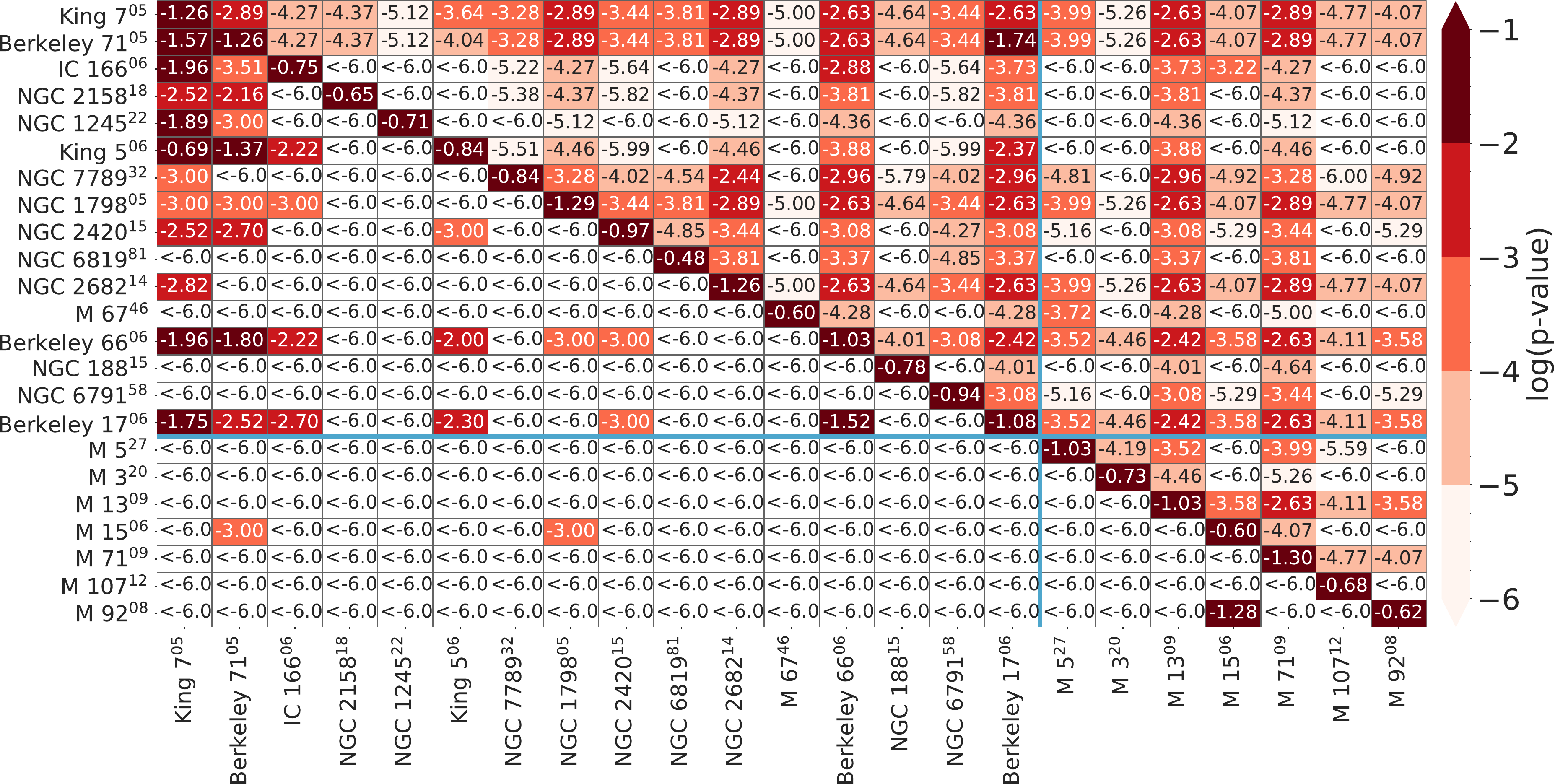}
	\caption{\label{fig: cramer} Median log p-values of 100 runs of the Cramer two-sample test for each pair of clusters. The elements below the main diagonal show the median value of 100 runs of the Cramer test, and the elements above the diagonal show the results for the K-S test. The cells are colored in shades of red as shown in the color bar. Inside each cell, we show the p-value of the test for each particular pair of clusters. Superscripts in the cluster names indicate the number of stars in the cluster. Two blue lines separate the objects into globular and open clusters.}
\end{figure*}

We note that passing the tests does not guarantee that the samples can be separed. The tests only evaluate whether two clusters can be drawn from the very same distribution. For example, two samples generated from two Gaussian distributions with the same center and different widths would be distinguishable by these tests but would not be fully separable by a clustering algorithm.

These tests are valuable for establishing the potentials and limitations of the unsupervised algorithms. From the Cramer test we conclude that we should not expect any algorithm to completely separate Berkeley 71 from King 7, IC 166 from King 7, NGC 1245 from King 7, King 5 from King 7, Berkeley 66 from King 7, Berkeley 17 from King 7, King 5 from Berkeley 71, Berkeley 66 from Berkeley 71, Berkeley 17 from Berkeley 66, and M15 from M92.

\begin{table}
	\centering
	\caption{\label{tab:best_elem} Second to the fifth column: the number of times that each element allows to distinguish some pair clusters in Figure \ref{fig: ks-test}. Sixth column: median {standard deviation} in the measure of abundances for all the stars in the sample.}
	\setlength{\tabcolsep}{.21em}
	\begin{tabular}{cccccc}
		Element & \multicolumn{4}{c}{N$_{best}$} & $\sigma_{[X/M]}$     \\ \hline
  &  glob-glob$^{i}$   &  glob-open$^{ii}$ &  open-open$^{iii}$ &  all comb. & \\ 
		C  &            6 &           20 &           43 &   69 &  0.18 \\
		N  &            2 &           27 &           32 &   61 &  0.20 \\
		Na &            7 &           23 &           23 &   53 &  0.84 \\
		Al &            1 &           10 &           11 &   22 &  0.62 \\
		Fe &            0 &           11 &            6 &   17 &  0.65$^{iv}$ \\
		O  &            0 &            7 &            8 &   15 &  0.12 \\
		K  &            4 &            7 &            4 &   15 &  0.60 \\
		Mg &            3 &            2 &            4 &    9 &  0.10 \\
		Ni &            2 &            4 &            0 &    6 &  0.61 \\
		Si &            2 &            1 &            2 &    5 &  0.12 \\
		P  &            1 &            0 &            2 &    3 &  0.72 \\
		S  &            0 &            0 &            1 &    1 &  0.18 \\
		Ca &            0 &            0 &            0 &    0 &  0.11 \\ \hline
		\multicolumn{6}{l}{\scriptsize{i: Pairs of globular clusters.}} \\
		\multicolumn{6}{l}{\scriptsize{ii: Pairs with one globular and one open cluster.}} \\
		\multicolumn{6}{l}{\scriptsize{iii: Pairs of open clusters.}} \\
		\multicolumn{6}{l}{\scriptsize{iv: \textbf{$\sigma_{[Fe/H]}$}.}} 
	\end{tabular}
\end{table} 

\section{Clustering algorithms}
\label{sec:clustering}

\quad We have applied eight different clustering algorithms for the classification of stellar clusters in chemical abundances space (\allabundances\footnote{The bracket notation for two given elements X and Y, [X/Y] is defined as
	\begin{equation}
	\label{eq:bracket}
	[\mathrm{X}/\mathrm{Y}]= \log_{10}{\left({\frac{N_{\mathrm{X}}}{N_{\mathrm{Y}}}}\right)_{\mathrm{star}}}-\log_{10}{\left({\frac{N_{\mathrm{X}}}{N_{\mathrm{Y}}}}\right)_{\odot}},
	\end{equation}
	where $N_{X}$ and $N_{Y}$ are the number of X and Y nuclei per unit volume, respectively.}) to determine how well they perform in separating known clusters. The algorithms are an affinity propagation \citep{affinity0}, agglomerative clustering \citep{agg}, DBSCAN  \citep{DBSCAN}, $K$-means \citep{macqueen}, mini-batch $K$-means \citep{minik}, spectral clustering  \citep{spectral}, Gaussian mixing models \citep{GMM}, and Bayesian Gaussian mixing models \citep{bayesian0, bayesian1}. These are all the nonhierarchical clustering algorithms available in the scikit-learn library \citep{scikit}. The complete description of these methods is beyond the scope of this paper; we refer to \citet{clustering_review} who provide a review of these algorithms and to the scikit-learn website, which contains the documentation for the library used in this work, \url{http://scikit-learn.org/stable/documentation.html}.

To evaluate the algorithm that is best suited for this task, we compared the algorithms through three different metrics (merit functions): homogeneity score, accuracy score, and v-measure score. When unsupervised clustering algorithms are run, the labels generated for the clusters can vary from one run to another. Even when the same objects are grouped together in both runs, their labels can differ. The v-measure score and homogeneity score are transparent to permutations of the labels, but the accuracy score needs all the clusters to be cross-matched. In this case, we matched each group of stars found by the unsupervised tool to the star cluster with the highest number of member stars inside the group, as was done in \cite{sanchez13} and \cite{k-means}. However, in this work when the number of clusters in the real dataset did not match the number of clusters in the predicted model, or when the objects in one group did not match any of the available clusters, we assigned the group to the cluster with the highest number of coincident objects, even when the cluster had previously been assigned to another group.

The accuracy score measures the number of coincidences between the real classification and the label assigned by for the clustering algorithm. It is a value from zero to one that represents the fraction of stars that are classified into the right cluster. The homogeneity score measures at which level the predicted clusters contain only data points that are members of one real cluster. The score value varies from zero to one, where 1 means that the clusters are perfectly homogeneous. The v-measure score is the harmonic mean between completeness and homogeneity. \cite{vscore} present a rigorous description of these merit functions.

For each of these algorithms, we performed an extensive optimization of their hyperparameters searching for the highest homogeneity score. The list of hyperparameters we tuned for each clustering method is shown in Table \ref{tab:hyper}, together with their best-fit values. The description of each of these parameters is given in the articles cited and also in the documentation of scikit-learn. The hyperparameters presented in Table \ref{tab:hyper} are labeled exactly as in the documentation of scikit-learn.

\begin{table}
	\caption{\label{tab:hyper} List of hyperparameters we explored for each algorithm. The last column shows the value of the parameter in the run with the highest homogeneity score.}
	\begin{tabular}{lll}
		Hyperparameters  & Tested values & best value \\ \hline
		\multicolumn{3}{c}{\textbf{Affinity Propagation}} \\
		affinity & euclidean &  euclidean \\
		convergence\_iter & 20 & 20 \\
		damping &  [0.5, 0.51, ..., 0.99] & 0.5 \\
		max\_iter & 200 & 200 \\ \hline
		\multicolumn{3}{c}{\textbf{Agglomerative Clustering}} \\
		affinity &  [manhattan, cosine, & \\
		 & euclidean] &  manhattan \\
		compute\_full\_tree & [False, True] & False \\
		linkage & [complete, average] & complete \\
		n\_clusters & [9, 10, ..., 26] & 26 \\ \hline
		\multicolumn{3}{c}{\textbf{Bayesian Gaussian Mixture}} \\
		covariance\_type & [tied, diag] & diag \\
		init\_params &  [kmeans, random] & kmeans \\
		n\_components &  [9, 10, ..., 26] & 20 \\
		n\_init & 5 & 5 \\
		random\_state$^\dagger$ &  [31, 43, ..., 473] &        283 \\
		warm\_start &      [False, True] &      False \\ \hline
		\multicolumn{3}{c}{\textbf{DBSCAN}} \\
		algorithm &         [ball\_tree, kd\_tree, & \\
		 & brute] &  ball\_tree \\
		eps &  [0.3, 0.31, ..., 1.49] &       0.58 \\
		leaf\_size & 10 &         10 \\
		metric & euclidean & euclidean \\
		min\_samples &                              [2, 3] &          2 \\ \hline
		\multicolumn{3}{c}{\textbf{Gaussian Mixture}} \\
		covariance\_type &  [spherical, diag] &  spherical \\
		init\_params &   [kmeans, random] &     kmeans \\
		max\_iter & 1000 & 1000 \\
		n\_components & [2, 3, ..., 26] & 20 \\
		random\_state$^\dagger$ &  [31, 43, ..., 473] &        283 \\ \hline
		\multicolumn{3}{c}{\textbf{$K$-means}} \\
		init &  [k-means++, random] &     random \\
		n\_clusters &  [9, 10, ..., 26] &         26 \\
		n\_init & 5 & 5 \\
		random\_state$^\dagger$ & [31, 43, ..., 473] & 43 \\ \hline
		\multicolumn{3}{c}{\textbf{Mini-batch $K$-means}} \\
		batch\_size & [10, 20, ..., 100] & 90 \\
		init &  [k-means++, random] &     random \\
		n\_clusters & [9, 10, ..., 26] & 25 \\
		random\_state$^\dagger$ & [31, 43, ..., 473] & 31 \\
		reassignment\_ratio &   [0.1, 0.01, 0.001] &        0.1 \\ \hline
		\multicolumn{3}{c}{\textbf{Spectral Clustering}} \\
		affinity &  [rbf, sigmoid, &  \\
		 & polynomial, poly] & rbf \\
		assign\_labels & [kmeans, discretize] &  discretize \\
		degree & [3, 4, 5] & 3 \\
		n\_clusters & [9, 10, ..., 26] &          17 \\
		n\_neighbors & [2, 5, 10] &           2 \\
		random\_state$^\dagger$ & [31, 43, ..., 473] &         383 \\ \hline
		\multicolumn{3}{l}{\scriptsize{$\dagger$: We have tried ten different random seeds, 31, 43,  98, 196, 283, 294, }} \\
		\multicolumn{3}{l}{\scriptsize{\quad 374, 383, 433 and 473.}} \\
	\end{tabular}
\end{table}

\subsection{Scalers}

\quad When high-dimensional data are used to perform clustering, it is essential to ensure all the dimensions are properly scaled. Many standard algorithms are availiable in the literature to achieve this kind of normalization. In this work, we have tested all the clustering algorithms with two different scaler algorithms (scikit-learn package; \citealt{scikit}) that are known as the standard scaler and the robust scaler. The standard scaler sets the mean value of all dimensions to zero and scales the variance of all dimensions to one. The robust scaler sets the median to zero and scales the data according to the range between the first and the third quantile, making it robust to outliers. 

When the two types of scaler were tested with the clustering algorithms, we found the highest homogeneity scores using the standard scaler for the affinity propagation, DBSCAN, and $K$-means algorithms. For the other five algorithms, the highest homogeneity score was found using the robust scaler.

\subsection{Dimensionality reduction}

\quad Dimensionality reduction is important not only to reduce computational cost but also to eliminate redundancy. Redundant dimensions can hamper the finding of clusters by the dilution of Euclidean distances when dimensions grow. This phenomenon is known as the curse of dimensionality, and it is discussed, for example, in \cite{dimensional}. We ran each clustering algorithm after dimensionality  reduction with a principal component analysis (PCA) \citep{PCA}, a linear discriminant analysis (LDA) \citep{LDA_0}, a independent component analysis (ICA) \citep{ICA}, a t-distributed stochastic neighbor embedding (t-SNE) \citep{TSNE}, and also without dimensionality reduction. 

For all clustering algorithms tested here, we found the highest homogeneity scores when the clustering algorithms were associated with the LDA dimensionality reduction algorithm. We varied the number of dimensions from 2 to 12 for all the dimensionality reduction tools. The number of dimensions that maximizes the homogeneity score differs among the algorithms. For the DBSCAN and spectral clustering, the highest homogeneity scores are found by projecting the data into two components. For agglomerative clustering and $K$-means, the best-suitable number of dimensions is 4. Five dimensions give the highest homogeneity score for the Gaussian mixture models and the mini-batch $K$-means. For affinity propagation, the best number of components is 6, and for the Bayesian Gaussian mixture models the best value is 7. {In machine-learning it is common practice to use the components that conserve a certain variance threshold, such as using the components that retain 90\% of the original variance in the dataset. This analysis shows that this approach to determine the best number of components can favor some clustering algorithms over others, because the best number of components depends on the clustering method.}

\section{Results of the clustering algorithms}
\label{sec:results}

\quad We tested the eight clustering algorithms listed in Sec. 5 by varying their hyperparameters and combining them with four different dimensionality (Sec. 5.2) reductions tools and two different scaling methods (Sec. 5.1). Figure \ref{fig:best_18} shows the best scores achieved for each of the clustering algorithms when the number of clusters is \nclusters, that is, the number of real stellar clusters {in this sample}. The algorithms are ordered by their best homogeneity score.

These results were obtained through working with the APOGEE uncalibrated elemental abundances, \allabundances, for \nstars\, stars in \nclusters\, clusters. Figure \ref{fig:conf} shows the confusion matrix of the clustering result with the highest homogeneity score, which was obtained with the mini-batch $K$-means. The confusion matrix compares the actual labels with the predicted labels. Each cell shows the ratio of the objects belonging to the class indicated in the vertical axis classified as belonging to the class in the horizontal axis. The main diagonal presents the true positives, namely when an object is classified in the cluster it belongs to, while the off-diagonal elements represent confusion between clusters (an object from one cluster is classified as belonging to another cluster). Of the six pairs that appear as indistinguishable in Figure \ref{fig: cramer}, four of them present a degree of confusion higher than 20\% in Figure \ref{fig:conf}. The pairs M3-M13, M67-NGC 6819, and NGC 188-NGC 6819, on the other hand, are slightly mixed in the classification but are not pointed out as problematic by the Cramer test. Because the test does not guarantee separability, these cases are to be expected.

\begin{figure}
	\centering
	\includegraphics[width=0.5\textwidth]{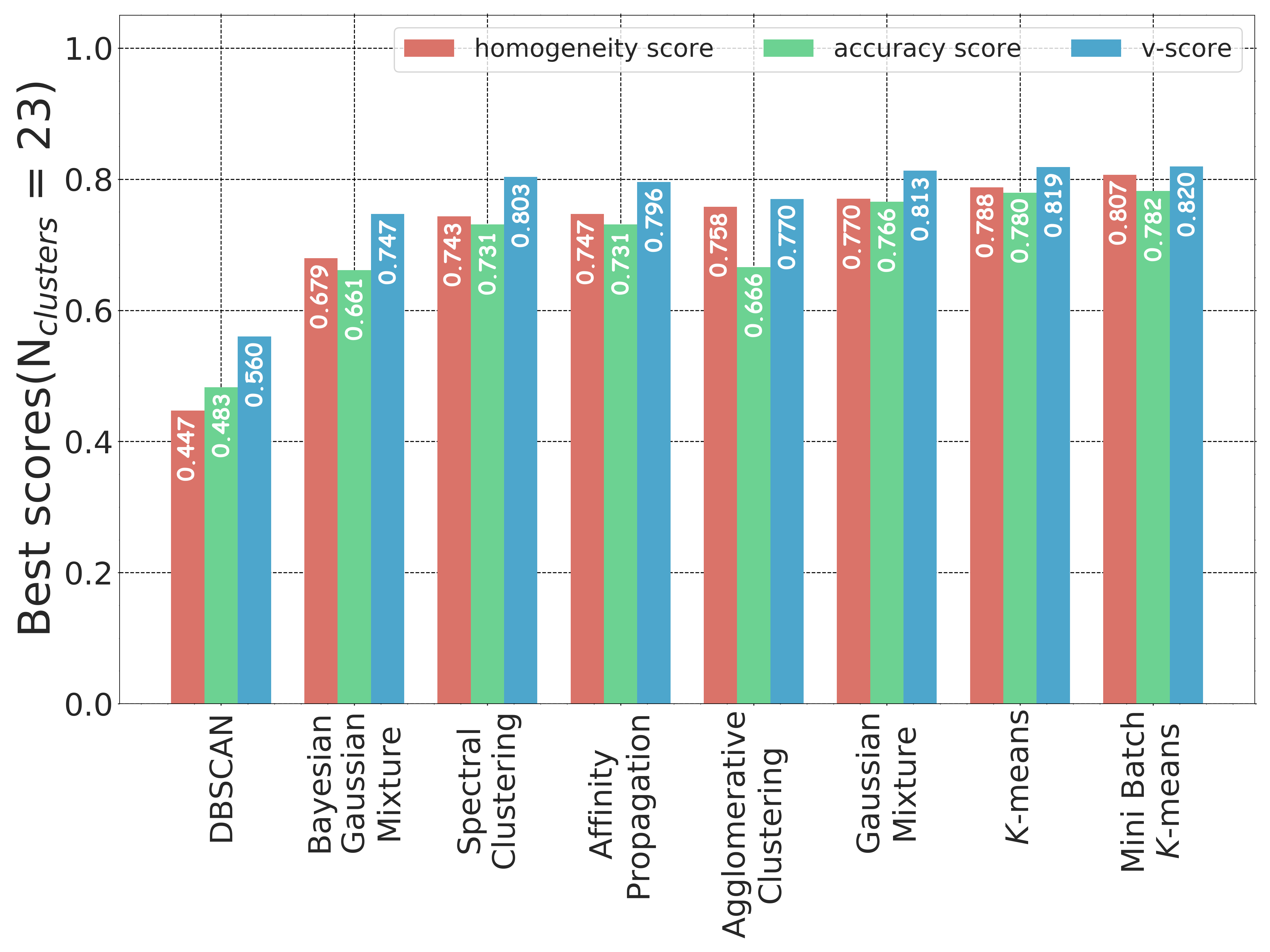}
	\caption{\label{fig:best_18} Best result for each of the algorithms tested in this work represented by bars. Different colors are used to distinguish among the metrics, as shown in the legend. In groups of three bars, the leftmost bar represents the homogeneity score, the bar in the middle represents the accuracy score, and the rightmost bar represents the v-measure score.}
\end{figure}

Figure \ref{fig:dr_nc} shows the variation in homogeneity score with the number of dimensions for all different dimensionality reduction algorithms, and for the two different scalers. {The highest homogeneity score is found} using LDA as dimensionality reduction. Moreover, LDA is almost unaffected by the choice of the scaler. The best result is found when we used LDA with nine dimensions: the homogeneity score is \besthomoscore. When no dimensionality reduction tools were used, we obtain the results shown in Table \ref{tab:nodr}. In this case, the highest homogeneity score is found for the Gaussian mixture model, 0.737, which is lower than we obtain using LDA but higher than we observe for the other dimensionality reduction tools.

\begin{table}
	\centering
	\caption{\label{tab:nodr} Highest scores found using clustering without dimensionality reduction.}
	\setlength{\tabcolsep}{.16667em}
	\begin{tabular}{lccc}
		Algorithm                    &  homogeneity & accuracy & v-measure \\ \hline
		Gaussian mixture             &   0.737 &  0.698 &     0.780 \\
		Mini-batch $K$-means         &   0.732 &  0.710 &     0.767 \\
		$K$-means                    &   0.725 &  0.673 &     0.754 \\
		Agglomerative clustering     &   0.651 &  0.624 &     0.687 \\
		Bayesian Gaussian mixture    &   0.642 &  0.610 &     0.745 \\
		Spectral clustering          &   0.610 &  0.548 &     0.621 \\
		DBSCAN                       &   0.362 &  0.436 &     0.503 \\
		Affinity propagation         &   0.288 &  0.346 &     0.433 \\
	\end{tabular}
\end{table}

The LDA is a supervised dimensionality reduction algorithm that uses knowledge of the classes to create a linear projection of the data that maximizes the separation among the classes. The projection could be determined using a few known clusters and might then be applied to any number of stars. When the search for stellar populations is made blindly, we could use the known star populations to determine the projection and apply this to the whole sample.

\begin{figure*}
	\centering
	\includegraphics[width=\textwidth]{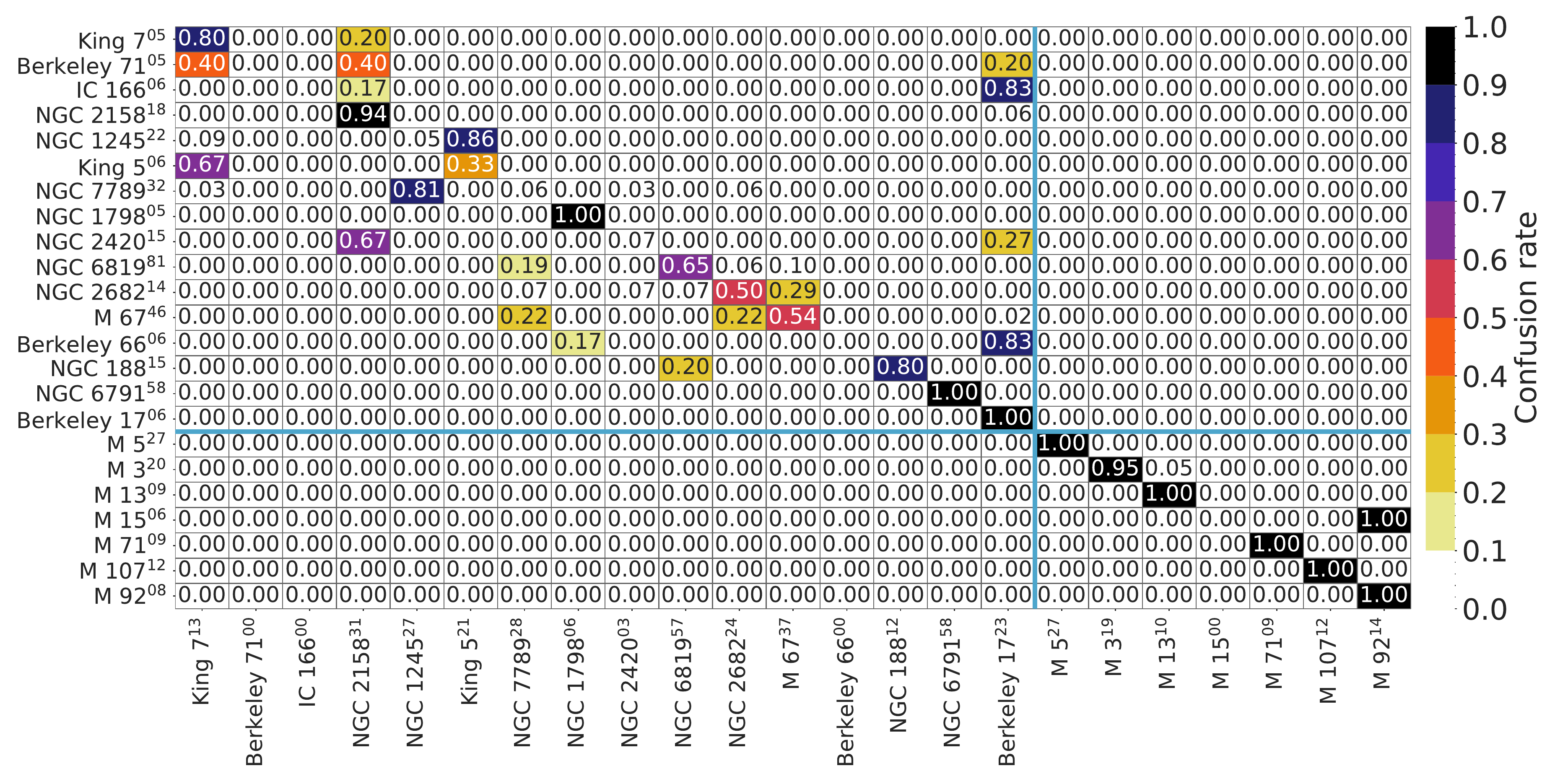}
	\caption{\label{fig:conf} Confusion matrix for our the best classification. The cells are color-coded according to the fraction of stars that are correctly classified as cluster members. The vertical axis corresponds to the real clusters, and the horizontal axis represents the clusters obtained with the mini-batch $K$-means. The main diagonal shows well-classified objects, and the cells out of the diagonal represent misclassifications.}
\end{figure*}

\begin{figure*}
	\centering
	\includegraphics[width=\textwidth]{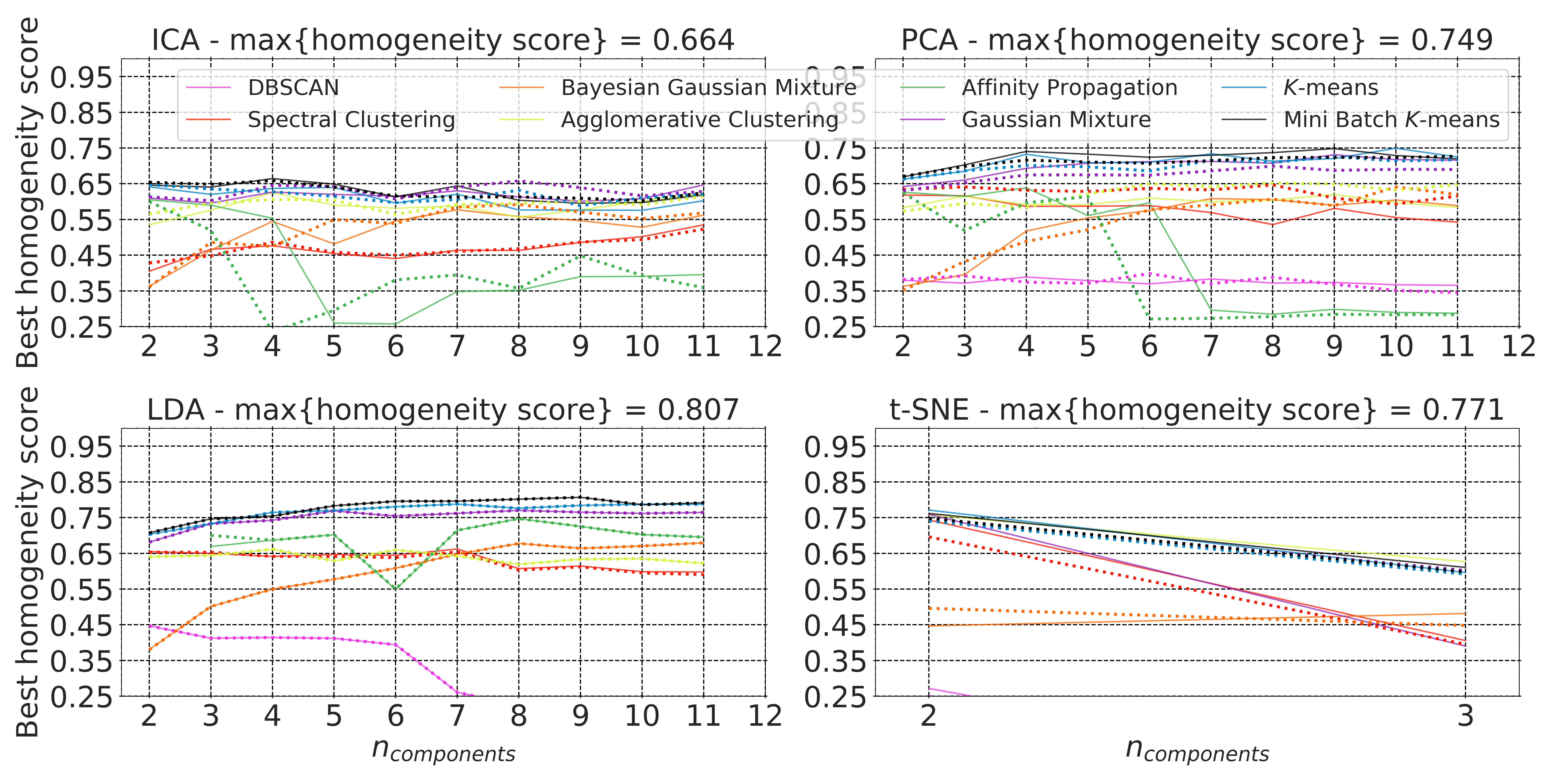}
	\caption{\label{fig:dr_nc} Best homogeneity score. This varies for each of the dimensionality reduction models with the variation of the number of components. The the labels at the top of each panel indicate the dimensionality reduction algorithm. Different colors are used for each clustering algorithm. Solid lines present the results obtained when we applied the robust scaler, and dotted lines show the results for the standard scaler.}
\end{figure*}

For comparison with \cite{Blanco-Cuaresma2015}, we present in Figure \ref{fig:best_all} the best scores without constraining the number of clusters to \nclusters. \cite{Blanco-Cuaresma2015} found the best result with the Mitschang algorithm\footnote{The Mitschang algorithm is not publicly available, therefore we were unable to test it against our sample.} \citep{mitschang}, with a homogeneity score of 0.86 and a v-measure score of 0.75. Our highest homogeneity score is 0.853, found using DBSCAN, with a v-measure score of 0.861. The quality of the data used by \citet{Blanco-Cuaresma2015} is arguably better than ours, in the sense that the spectral resolution of their spectra is higher than in APOGEE, and the authors used chemical abundances for 17 elements, while we used 13. {In addition, we did not use the same clusters as they did, which can bias the comparison.} However, {assuming that the overall distribution of chemical abundances of both samples are comparable}, our results show that APOGEE data are capable of yielding results {at a similar level as those obtained by \citet{Blanco-Cuaresma2015} from optical spectroscopy}.

\begin{figure}
	\centering
	\includegraphics[width=0.5\textwidth]{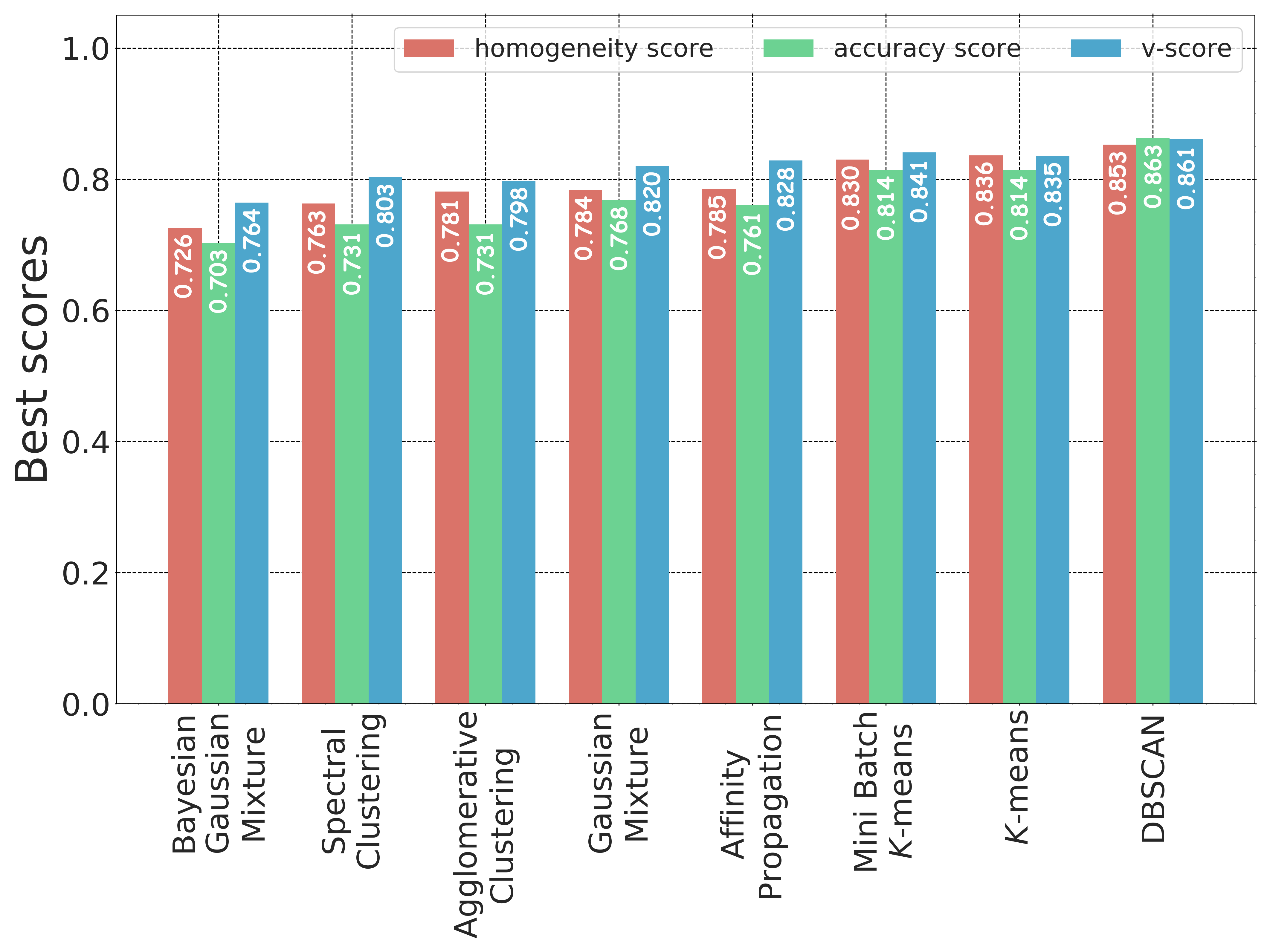}
	\caption{\label{fig:best_all} Best result for each of the algorithms we tested without constraining the number of clusters, represented by bars. Different colors are used to distinguish among the metrics, as shown in the legend. In groups of three bars, the leftmost bar represents the homogeneity score, the bar in the middle represents the accuracy score, and the rightmost bar represents the v-measure score. The actual values are given in each bar.}
\end{figure}

\section{Conclusions}

\quad We have explored the application of unsupervised clustering on the APOGEE survey to chemically separate star clusters from each other. We statistically tested the feasibility of the task and concluded that it cannot be accomplished perfectly because of the intrinsic overlap of the clusters in chemical space. Eight different clustering algorithms were combined with four dimensionality reduction techniques and two scaling approaches. We have shown that the highest homogeneity score obtained from the clustering process is consistent with the expectations from the Cramer test.

The Kolmogorov-Smirnov test allowed us to identify the chemical elements that are the most sensitive for the identification of stellar populations. Seven elements are sufficient to distinguish 90\% of the pairs of clusters. Table \ref{tab:best_elem} shows that the best set of elements depends on the type of cluster. For example, some elements are more relevant for separation globular clusters from each other (Fe and K), while other elements are more suitable for distinguishing open from globular clusters (C, Al, and Na). This information is highly relevant when the decision needs to be made whether to allocate computational and observational resources to improve the precision of abundance measurements or to expand the number of elemental species to be measured. {We stress that this information is restricted to the set of elements explored here (\allabundances) and the stellar populations we studied.}

The Cramer test suggests that there are six pairs of indistinguishable clusters in our sample, M2-M5, M2-M13, M15-M92, NGC 2158-NGC 2420, NGC 2158-Pleiades, and NGC 2420-Pleiades. This does not mean the other pairs are fully separable, because different but overlapping distributions can be distinguishable and not separable. On the other hand, it does not mean the that indistinguishable cluster pairs are intrinsically identical; it only guarantees that the particular samples of stars and chemical elements that we have studied are indistinguishable. It is possible that the same stars measured with higher precision, the use a different set of elements, or the availability of larger samples of stars in the analysis, could lead to a different conclusion.

We have tried to separate the cluster members using unsupervised classification algorithms. These reached a maximum homogeneity score of 0.85, where the confusion is primarily associated with pairs that were marked as indistinguishable by the Cramer test. The best result was found using mini-batch $K$-means, but $K$-means and Gaussian mixture models give a very similar performance. Gaussian mixture models offer a more elegant solution, providing not only the classification of the stars, but also the probability of belonging to other groups. Moreover, the mixture of Gaussian functions can generate decision boundaries that can adapt better to the intrinsic form of the clusters, while $K$-means like algorithms assume hyperspherical boundaries{, when Euclidean metric is applied}. However, in larger samples where the computational cost can be a constraint, mini-batch $K$-means would be preferable.

%We have tried to determine the number of clusters automatically. None of the criteria were able to recover the exact number of clusters. However, the BIC and the gap statistics give results consistent with the number of clusters we expect due to the indistinguishability of some pairs of clusters.

In summary, we have tested the limits of the distinguishability of the stellar clusters with APOGEE data and explored the performance of various clustering algorithm to reach these limits. In this sense, we have shown that the chemical identification of stellar populations is limited by the available data. We have slightly improved the results of the classification compared with previous attempts. With the chemical information provided by APOGEE, it is not possible to completely distinguish all the stellar clusters from each other. Even though we are not able to completely separate the clusters from each other, the primary sources of confusion are clusters with similar ages, as is the case for the three globular clusters M2, M3 and M13, and the pair of open clusters NGC 2158 and NGC 2420. {On the other hand, old metal-poor globular clusters were much easier to distinguish, as could be naturally expected because their chemical patterns are distinct Therefore, strong chemical tagging might yield better results on these populations.}

In a recent study, \cite{Ness2018} demonstrated the existence of stars with almost identical chemical composition, but with a different galactic origin, which adds to the limitations presented in this work. However, our results indicate that if chemical tagging is not possible to the level of star clusters, the existing clustering algorithms can blindly identify stellar populations with similar ages and chemical distributions in APOGEE data (but not the exact clusters).

Traditionally, the chemical distributions of stellar populations are either used to distinguish large-scale Galactic components, such as the thin and thick disk, or to identify populations at the star cluster level. The results found in this paper add to what has been found by \citet{Blanco-Cuaresma2015} and \citet{Ness2018} in demonstrating the difficulties for blind searches to identify star clusters from their chemical abundances.

There is a possibility of improving the spectroscopic data or including radial velocities or proper motions in the cluster identification, which would proceed the process to a chemical and kinematic tagging {as in \citealt{Chen_2018}}. Alternatively, chemical tagging might be thought of as an intermediate level between the Galactic components and the stellar clusters. If we identify many stars that are chemically very similar to those in a cluster, we expect them to share a similar chemical evolution history with the cluster, even if the two sets are spatially unrelated.

%For many decades the star clusters were the main source of well-measured distances and ages through the Galactic disk. In the era of Gaia \citep{gaiamission}, astronomers could lose the interest for this kind of object, since distances and ages now can be measured with much higher precision. However, we argue that by relating field stars to families of star clusters we could extend the knowledge from the clusters to the stars in the Galaxy.

\appendix

\section{Main results with calibrated abundances}
\label{sec:ap}
In order to enrich the description of the analysis while preserving a comprehensive flow of ideas in the article, we reserved the presentation of the results for the calibrated sample in this appendix. Here we reproduce the main relevant results we presented in the analysis of the article to demonstrate that the choice for uncalibrated abundances only quantitatively affects the results. The conclusions presented in the main section of the article remain the same. Here the initial sample was the same as the uncalibrated sample, and all steps were applied in exactly the same manner. We therefore refer to the main sections for more details in the analysis.

Although we have started with the same dataset for the calibrated analysis, the cleaning process was more severe in this case: fewer star per cluster were kept after the sigma clipping. Because we established a minimum threshold of five stars per clusters, here the analysis is based on fewer clusters. In Figure \ref{fig: C_calib} the clusters King 7, M107, M15 and M92 were excluded from the analysis with calibrated abundances. We also excluded the abundances for Al and Na because we lack of measurements for 11\% and 18\% of the stars in the sample, respectively.

In Figures \ref{fig: C_calib}, \ref{fig: ks-test_calib}, and \ref{fig: cramer_calib} we show the results of the K-S test and the Cramer test for all clusters in the calibrated sample. The degree of confusion among the clusters is even more severe than in the uncalibrated abundances. The indistinguishability of more pairs of clusters results in a poorer performance of the algorithms in separating the clusters, as shown in Figures \ref{fig:conf_calib} and \ref{fig:best_all_calib}.

This small section supports the focus on the uncalibrated for the rest of the analysis. The same conclusions as presented for the uncalibrated sample apply to the calibrated sample. However, working with uncalibrated data is a more conservative approach in our hypothesis of cluster indistinguishability because this characteristic is more even more obvious in the calibrated data and is probably aggravated by the calibration.

\begin{figure*}
	\centering
	\includegraphics[width=\textwidth]{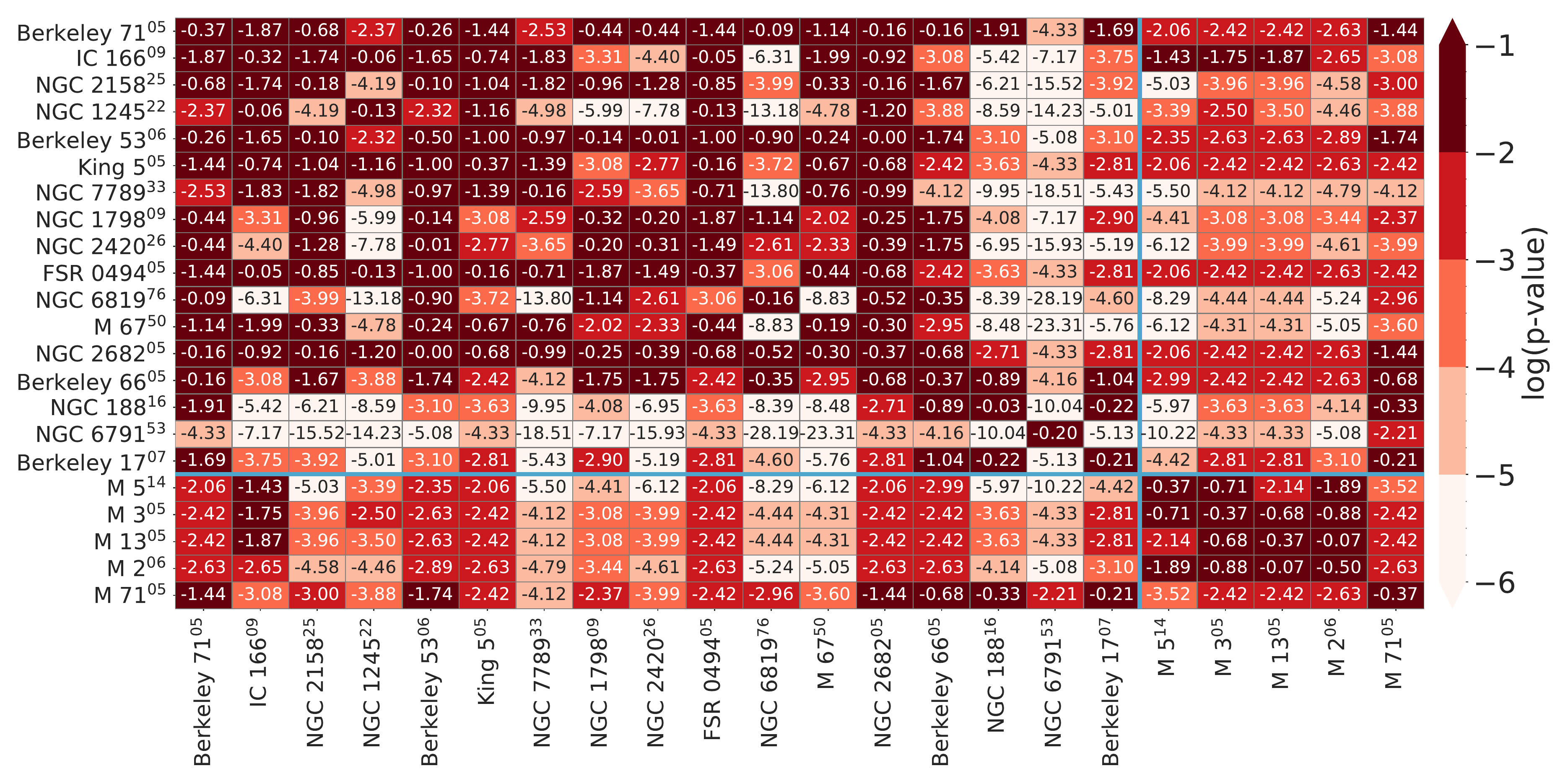}
	\caption{\label{fig: C_calib} Same as Figure \ref{fig: C} for the calibrated abundances. Logarithm of the p-values of the K-S two-sample test for the carbon distributions of each pair of clusters. The cells are colored in five shades of red, from light to dark in logarithmic scale. Inside each cell, we show the p-value of the test for that particular pair of clusters. Superscripts in cluster names indicate the number of stars in the cluster. The main diagonal shows the median p-value of 1000 tests resulting from randomly dividing the cluster into two subsamples of nearly equal sizes. Two blue lines separate the objects into globular clusters (lower right) and open clusters (upper left). Dark red represents values greater than 0.01: we cannot reject the null hypothesis for these cases.}
\end{figure*}

\begin{figure*}
	\centering
	\includegraphics[width=0.99\textwidth]{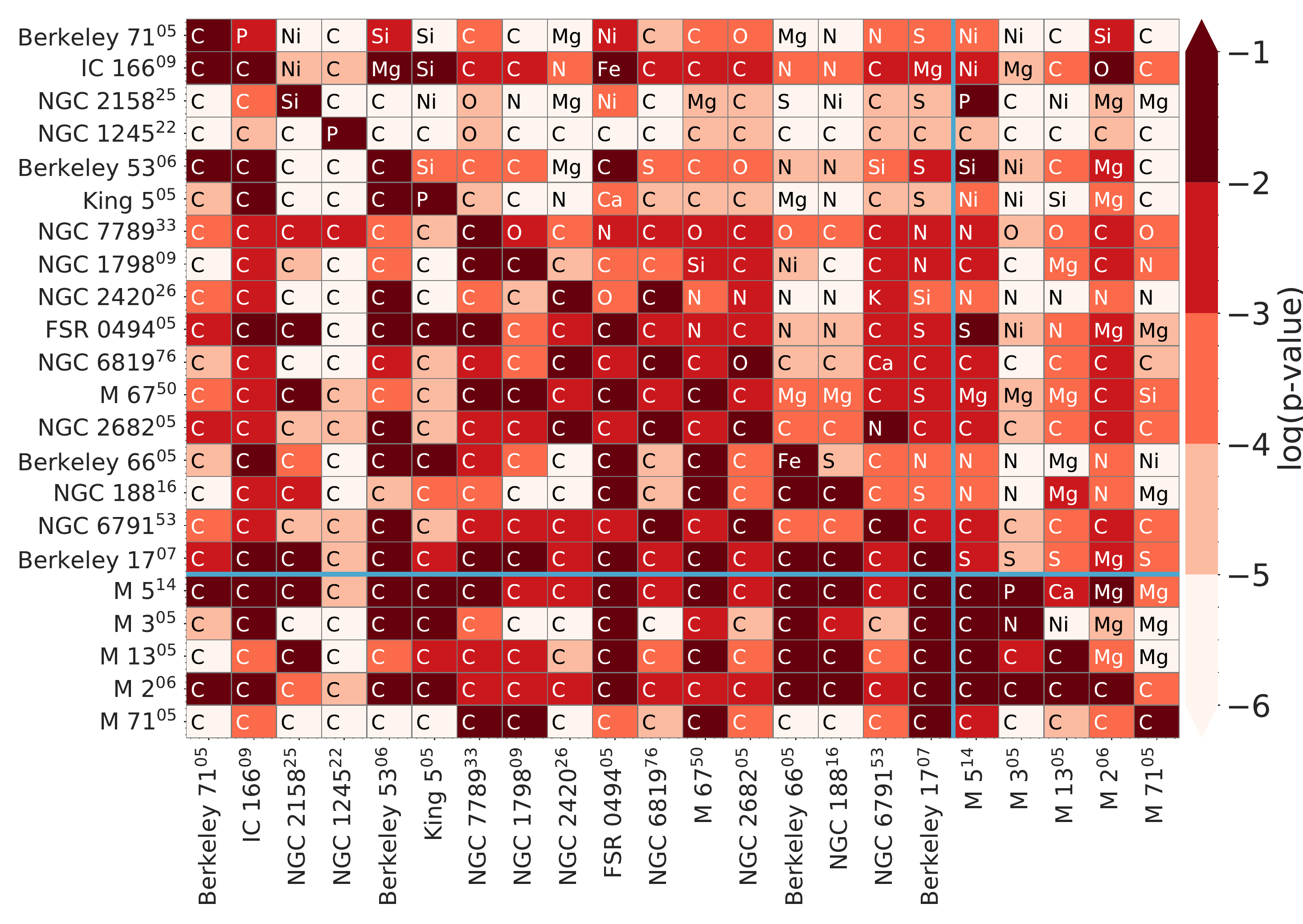}
	\caption{\label{fig: ks-test_calib} Same as Figure \ref{fig: ks-test} for the calibrated abundances. Logarithm of the p-values of the K-S two-sample test for each pair of clusters. The elements of the main diagonal and above the main diagonal represent the lowest p-value found among all the elements, while the elements below the main diagonal represent the p-value for carbon. The cells are colored in shades of red as shown in the color bar. Red colors are saturated at 0.01. Superscripts in the cluster names indicate the number of stars in the cluster. At the center of each cell, we show the element for which the p-value is calculated. Two blue lines separate the objects into globular and open clusters.}
\end{figure*}

\begin{figure*}
	\centering
	\includegraphics[width=\textwidth]{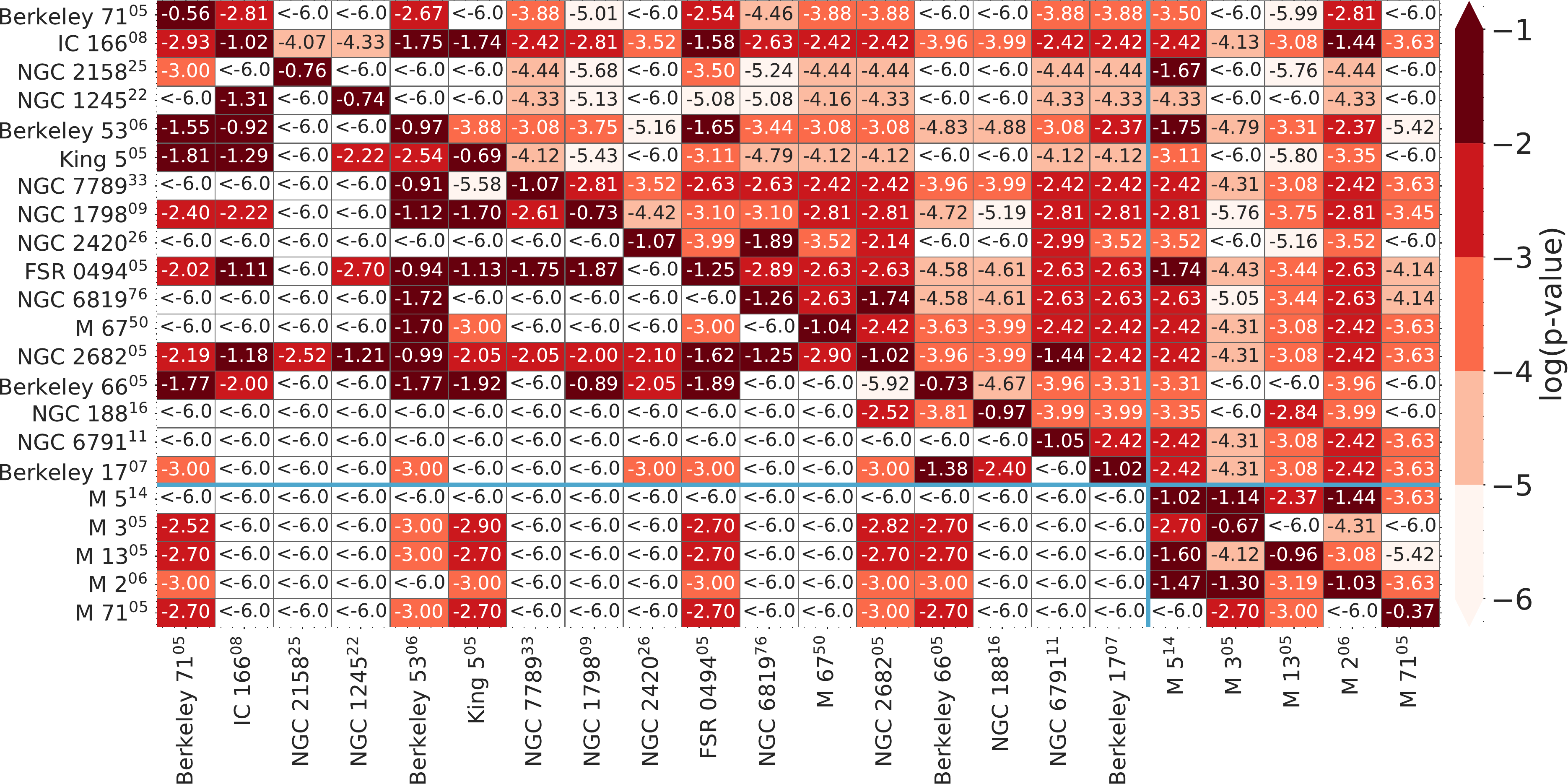}
	\caption{\label{fig: cramer_calib} Same as Figure \ref{fig: cramer} for the calibrated abundances. Median log p-values of 100 runs of the Cramer two-sample test for each pair of clusters. The elements below the main diagonal show the median value of 100 runs of the Cramer test, and the elements above the diagonal show the results for the K-S test. The cells are colored in shades of red as shown in the color bar. Inside each cell, we show the p-value of the test for each particular pair of clusters. Superscripts in the cluster names indicate the number of stars in the cluster. Two blue lines separate the objects into globular and open clusters.}
\end{figure*}

\begin{figure*}
	\centering
	\includegraphics[width=\textwidth]{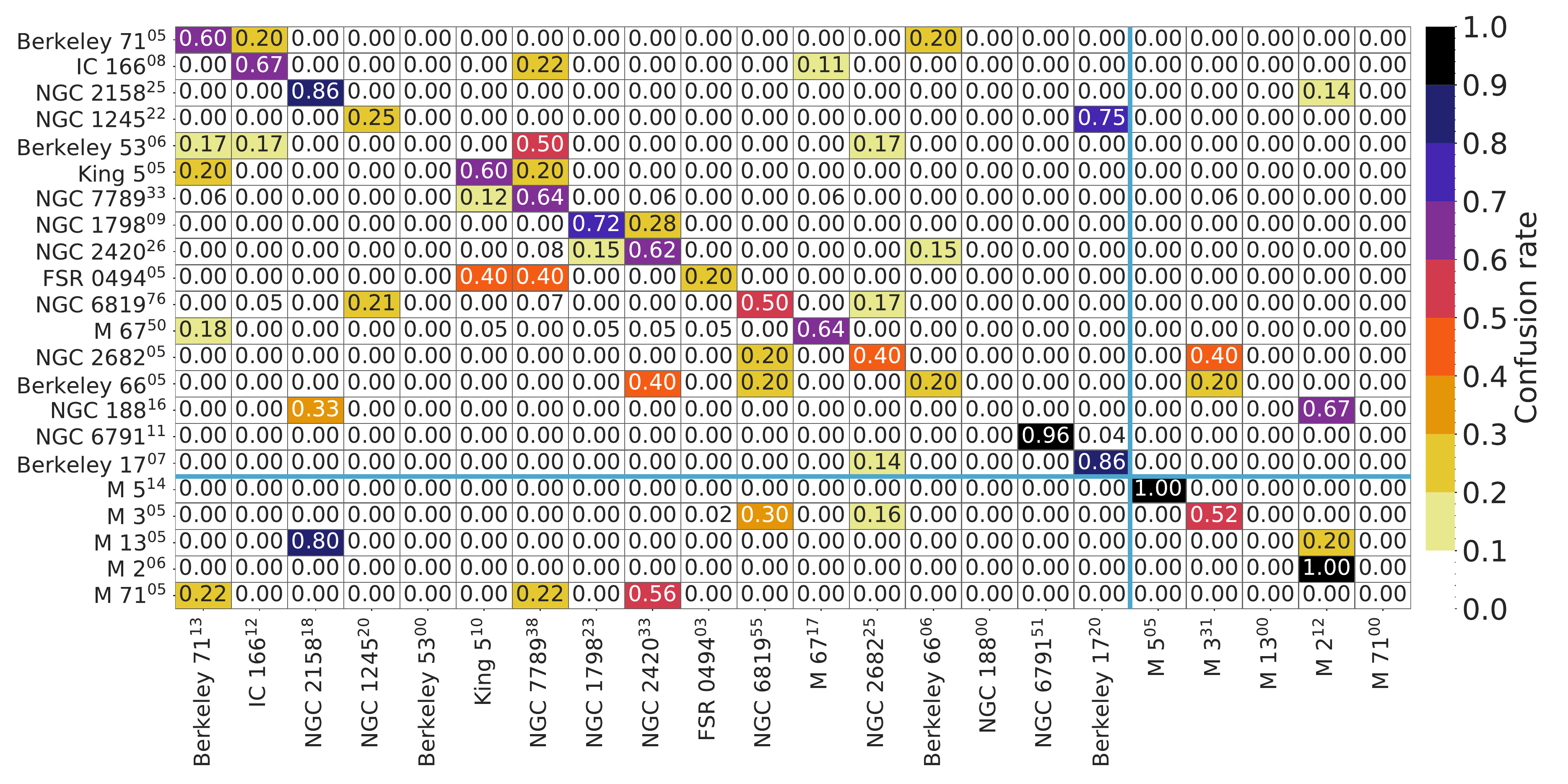}
	\caption{\label{fig:conf_calib} Same as Figure \ref{fig:conf} for the calibrated abundances. The confusion matrix for our the best classification. The cells are color-coded according to the fraction of stars that is correctly classified as cluster members. The vertical axis corresponds to the real clusters, and the horizontal axis represents the clusters obtained with the mini-batch $K$-means. The main diagonal shows well-classified objects, and the cells out of the diagonal represent misclassifications.}
\end{figure*}

\begin{figure}
	\centering
	\includegraphics[width=0.5\textwidth]{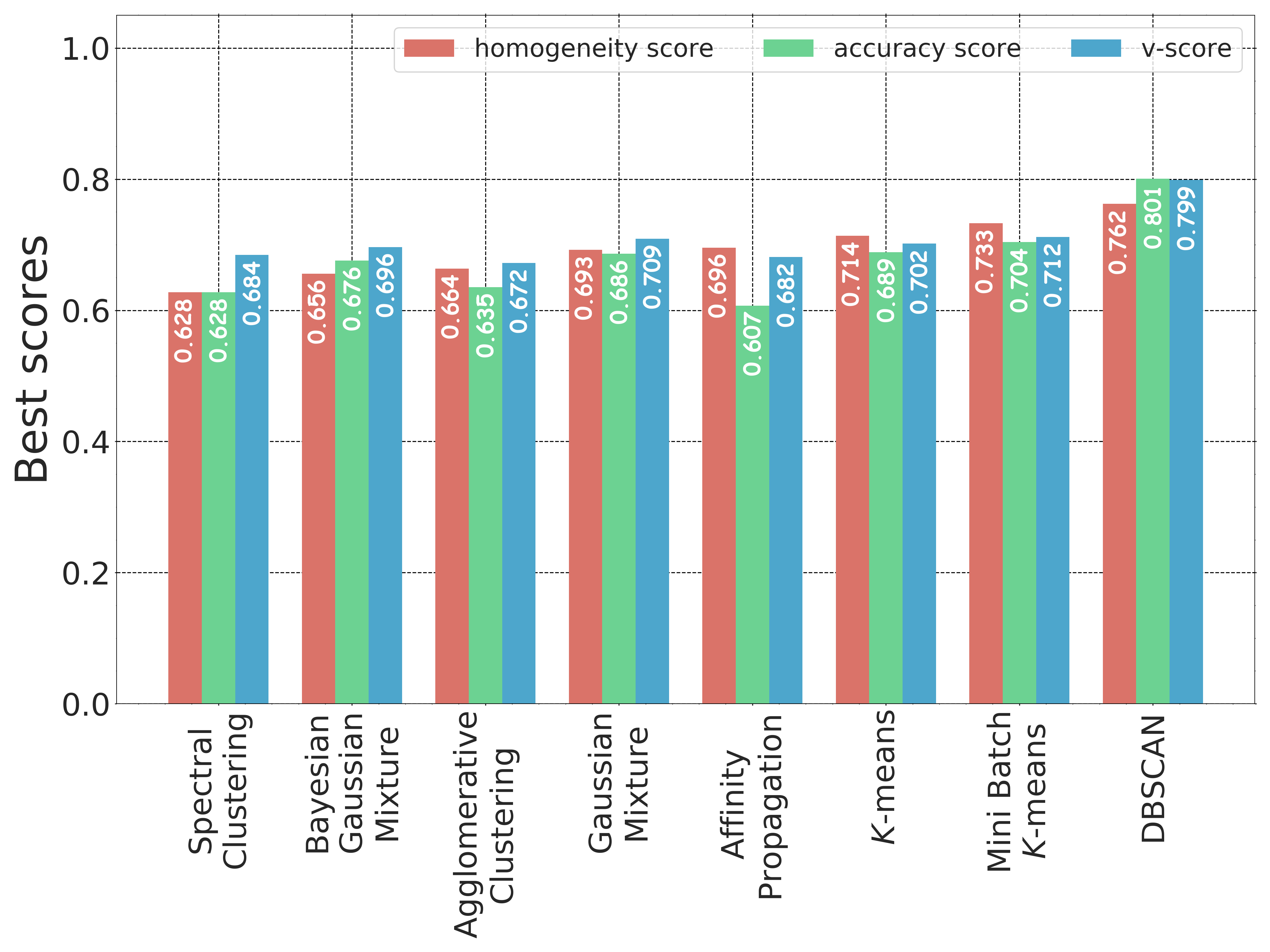}
	\caption{\label{fig:best_all_calib} Same as Figure \ref{fig:best_all} for the calibrated abundances. The bars represent the best result for each of the algorithms we tested in this work without constraining the number of clusters. Different colors are used to distinguish the metrics, as shown in the legend. In groups of three bars, the leftmost represents homogeneity score, the bar in the middle represents the accuracy score, and the rightmost bar represents the v-measure score. The actual values are given in each bar.}
\end{figure}

\begin{acknowledgements}
	
We acknowledge financial support through grants AYA2014-56359-P, AYA2017-86389-P, and AYA2016-79724-C4-2-P (MINECO/FEDER). The research that led to this article was partially funded by the Brazilian National Research Council (CNPq) through scholarship of the CSF program. CAP is thankful to the Spanish Government for funding for his research through grant AYA2014-56359-P. Funding for SDSS-III has been provided by the Alfred P. Sloan Foundation, the Participating Institutions, the National Science Foundation, and the U.S. Department of Energy Office of Science. The SDSS-III website is http://www.sdss3.org/.

SDSS-III is managed by the Astrophysical Research Consortium for the Participating Institutions of the SDSS-III Collaboration including the University of Arizona, the Brazilian Participation Group, Brookhaven National Laboratory, Carnegie Mellon University, University of Florida, the French Participation Group, the German Participation Group, Harvard University, the Instituto de Astrofisica de Canarias, the Michigan State/Notre Dame/JINA Participation Group, Johns Hopkins University, Lawrence Berkeley National Laboratory, Max Planck Institute for Astrophysics, Max Planck Institute for Extraterrestrial Physics, New Mexico State University, New York University, Ohio State University, Pennsylvania State University, University of Portsmouth, Princeton University, the Spanish Participation Group, University of Tokyo, University of Utah, Vanderbilt University, University of Virginia, University of Washington, and Yale University.

\end{acknowledgements}

\bibliographystyle{aa}
\bibliography{bibtex}

\end{document}